\documentclass{revtex4}
\usepackage{amssymb,amsfonts,amsmath,amsthm,mathtext,braket, mathtools, relsize}

\begin{document}

\title{On characteristics of mixed unitary channels being additive or multiplicative with respect to taking tensor products
}

\author{\firstname{G.G.}~\surname{Amosov}}
\email[E-mail: ]{gramos@mi-ras.ru}

\address{Steklov Mathematical Institute of Russian Academy of Sciences, ul. Gubkina 8, Moscow 119991, Russia}


\begin{abstract} 
We study mixed unitary channels generated by finite subgroups of the group of all unitary operators in a Hilbert space. Based on the majorization theory we introduce techniques allowing to calculate different characteristics of output states of channels. A class of channels has been allocated for which the use of entangled states doesn't give any advantage under taking supremum and infimum for output characteristics of channels. In particular, $l_p$-norms are multiplicative and the minimal entropy is additive with respect to taking tensor products of channels. As an important application of the obtained results the classical capacity of channel is calculated in the evident form. We compare our techniques with the informational characteristics of Boson quantum channels.
\end{abstract}


\maketitle

\section {Introduction}

The classical capacity of a quantum channel is a numerical characteristic showing how many quantum states can be used for encoding in order to transmit them over n copies of the channel so that the result can be decoded with any degree of accuracy at $n\to +\infty $. The main obstacle to calculating the classical capacity is the presence of entangled states of a composite quantum system \cite {Holevo}. One of the possible approaches to solve this problem is the search for quantum channels that behave like classical ones. That is, such channels for which the use of entangled states does not give any information gain. Until recently, only several important types of channels with this property are known \cite{Shor, King1, King2}. On the other hand, it is known that this property is not fulfilled for all quantum channels in general \cite{Hast}. Recently, the theory of majorization was used to solve such a problem \cite{urRehman, Siu}, which made it possible to achieve some progress \cite{Amo, Amo2}.

Calculating the capacity of a quantum channel is one of the most important open problems of quantum information theory. In the first attempts to solve this problem, it was proposed to investigate a wider class of output characteristics of quantum channels that depend on the eigenvalues of quantum states at the channel output. In particular, such characteristics are $l_p$-norms of channels determined through the sum of degrees of eigenvalues for the output states of the channel. This approach leads to the problem of multiplicativity of channel norms with respect to the operation of taking the tensor product \cite{Amo3, Amo4}. As it turned out, the multiplicativity property of norms, which is fulfilled in the absence of a gain from the use of entangled states when calculating the supremum, is also not always fulfilled \cite{Werner}. Thus, there is a broader program of research on the behavior of various information characteristics of channels with respect to the operation of taking a tensor product.

In search of a counterexample to the additivity property of the minimum output entropy of the channel with respect to taking the tensor product M.B. Hastings \cite {Hast} considered mixed unitary channels  whose action is a convex sum of actions performed by unitary operators. Unitary operators are selected randomly and such a measure is introduced on the set of all channels, which allows us to assert that the measure of channels with violation of the additivity property is greater than zero. The assumption that not all unitary operators are used in the definition of a channel, but only those belonging to some finite group, should significantly improve the properties of such a class of channels. Moreover, tensor products of such channels can be associated with the scheme of quantum computing, in which unitary operators from the group act as gates. Using this correspondence, we move from the operation of taking tensor products to the operation of sequential calculations of gates.

Developing the ideas proposed in \cite{Amo, Amo2} we study mixed unitary channels generated by unitary group.  Second section is devoted to basic definitions, notations and the basic known results in majorization theory.  In Section 3 the necessary technical results are obtained. At first Introducing the definition of majorization for a quantum channel we exploit the majorization theory \cite{Karamata} in order to establish how the majorization property is affected by a composition of two channels. Next, we introduce marginal distributions for mixed unitary channels and formulate the majorization condition associated with them. Then we get the properties of channels for which the majorization condition holds true. Section 4 is devoted to the concrete example in which the unitary group is obtained by the extension of Abelian finite group. We prove that if the majorization condition is satisfied for a channel, then the infimum of  a trace of any convex function from the output state of channel  is attained on the marginal distribution. This result allows us to show that the $l_p$-norms of the channel are multiplicative and the minimal output entropy is additive with respect to taking tensor product. The classical capacity is obtained for the channels belonging to the specified class. Section 5 is devoted to the discussion of the connection of the introduced techniques to estimating the Bosonic channels. The last section is for conclusion remarks.

\section{Preliminaries}

Throughout this paper we use the following notations \cite{Holevo, Bhatia}:

$H$ a separable Hilbert space;

${\rm I}_H$ the identity operator in $H$;

$B(H)$ the algebra of all bounded operators in $H$;

$\mathfrak {S}(H)$ the convex set of all quantum states (positive unit trace operators) $\rho $ in $H$;

$\Phi :\mathfrak {S}(H)\to \mathfrak {S}(H)$ a quantum channel (complete positive linear map preserving a trace) in $H$;

${\rm Id}_H:\mathfrak {S}(H)\to \mathfrak {S}(H)$ the identity channel in $H$;

Given two operators $x\in B(H),\ y\in B(K)$ we write $x\cong y$ iff there exists a unitary operator $W:H\to K$ such that
$WxW^*=y$.

Given a probability distribution $P=(p_j)$ we denote $P^{\downarrow}=(p_j^{\downarrow})$ the same distribution renumbered in the descending order:
$$
p^{\downarrow}_1\ge p^{\downarrow}_2\ge p^{\downarrow}_3\ge \dots
$$

Given two probability distributions $P=(p_j,\ 1\le j\le n)$ and $Q=(q_j,\ 1\le j\le n)$ we shall use a notation $P\prec Q$ if
$$
\sum \limits _{j=1}^kp_j^{\downarrow}\le \sum \limits _{j=1}^kq_j^{\downarrow},\ 1\le k\le n.
$$
Our approach is based upon the known result below.

{\bf The Karamata's theorem \cite {Karamata, Bhatia}.} {\it The following statements are equivalent:
\begin{itemize}
\item $P\prec Q$;
\item $\sum \limits _{j=1}^nF(p_j)\le  \sum \limits _{j=1}^nF(q_j)$ for all convex functions $F$.
\end{itemize}
}

Let $\mathcal U$ and $\mathcal T$ be a group and its subgroup. Given $U\in {\mathcal U}$ the set $\{UT,\ T\in {\mathcal U}\}$ is said to be a left coset of $\mathcal T$ in $\mathcal U$ and denoted by $[U]$. If $\mathcal T$ is a normal subgroup of $\mathcal U$ ($UTU^{-1}\in {\mathcal T}$ for all $U\in {\mathcal U},\ T\in {\mathcal T}$) the set of all cosets $[U],\ U\in {\mathcal U}$ is a group with respect to a multiplication $[U][V]=[UV],\ U,V\in \mathcal {U}$ known as a quotient group ${\mathcal U/T}$ \cite{Hall}. 

\section{Estimating eigenvalues of output states} 

To develop our approach we need the following definition.

{\bf Definition 1.}  {\it A quantum channel $\Phi :\mathfrak {S}(H)\to \mathfrak {S}(H)$ is said to be majorized by the probability distribution $P=P^{\downarrow }=\{p_j,\ 1\le j\le m\},\ m\le dimH$  iff
$$
\sum \limits _{j=1}^k\braket {e_j,\Phi (\ket {f}\bra {f})e_j}\le \sum \limits _{j=1}^kp_j,\ 1\le k\le m,
$$
for any choice of pairwise orthogonal unit vectors $(e_j)$ and a unit vector $f$ in $H$. If $\Phi $ is majorized by $P$ we shall write $\Phi \prec P$.}

Given two probability distributions $P=P^{\downarrow}=\{p_j,\ 1\le j\le m\}$ and $Q=Q^{\downarrow}=\{q_k,\ 1\le k\le n\}$ we denote $P\star Q=\{(p_jq_k)^{\downarrow},\ 1\le j\le m,\ 1\le k\le n\}$.

{\bf Proposition 1.} {\it Suppose that $\Phi \prec P$, $\Psi \prec Q$ and $dimH\ge mn$. Then, 
$$
\Phi \circ \Psi \prec P\star Q.
$$
} 

Proof.

Take a unit vector $f\in H$ and apply $\Psi $ to the corresponding projection. Let us consider a spectral decomposition
$$
\Psi (\ket {f}\bra {f})=\sum \limits _{j=1}^N\lambda _j^f\ket {f_j}\bra {f_j},
$$
where  eigenvalues $\lambda _j^f=\lambda _j^{f\downarrow}$ are arranged in the descending order.
Then,
$n$ first eigenvalues $\lambda _j^f$  satisfy the relation 
\begin{equation}\label{formula1}
\sum \limits _{j=1}^k\lambda _j^f\le \sum \limits _{j=1}^kq_j,\ 1\le k\le n.
\end{equation}
Applying $\Phi $ to each $\ket {f_j}\bra {f_j}$ we obtain
\begin{equation}\label{formula2}
\sum \limits _{s=1}^k\braket {e_s,\Phi (\ket {f_j}\bra {f_j})e_s}\le \sum \limits _{s=1}^kp_s,\ 1\le k\le m,
\end{equation}
for any choice of $m$ pairwise orthogonal vectors $(e_s)$. Consider the composition
\begin{equation}\label{formula3}
\sum \limits _{s=1}^k\braket {e_s,[\Phi \circ \Psi ](\ket {f}\bra {f})e_s}=\sum \limits _{s=1}^k\sum \limits _{j=1}^N\lambda _j^f\braket {e_s,\Phi (\ket {f_j}\bra {f_j})e_s}
\end{equation}
Combining estimations (\ref {formula1}) and (\ref {formula2}) we obtain that (\ref {formula3}) can be estimated from above  as we need.

$\Box $

Let $\mathcal U$ be a subgroup of ${\mathcal U}(H)$.  Consider a mixed unitary channel 
\begin{equation}\label{kanal}
\Phi (\rho )=\sum \limits _{U\in {\mathcal U}}\pi _UU\rho U^*,\ \rho \in \mathfrak {S}(H),
\end{equation}
$\pi _U\ge 0,\ \sum \limits _{U\in {\mathcal U}}\pi _U=1$.
Put
\begin{equation}\label{exp}
{\mathbb E}(x)=\frac {1}{|{\mathcal U}|}\sum \limits _{U\in {\mathcal U}}UxU^*,\ x\in B(H),
\end{equation}
then ${\mathbb E}$ is the conditional expectation onto the algebra $\mathcal A$ consisting of fixed elements with respect to the actions $x\to UxU^*,\ U\in {\mathcal U}$. 

{\bf Definition 2.} {\it We shall call that $\mathcal A$ has a codimension $n$ if
\begin{equation}\label{dim}
{\mathbb E}(\rho)\cong \frac {1}{n}{\rm I_n}\otimes \sigma,\ \rho \in \mathfrak {S}(H),
\end{equation}
where ${\rm I}_n$ is the identity operator in a Hilbert space of dimension $n<+\infty$ and $\sigma \in {\mathfrak S}(K)$ for some Hilbert space $K$. In particular the second multiplier on the right hand side of (\ref {dim}) can be omitted ($dimK=1$). 
}

{\bf Remark 1.} {\it The condition (\ref {dim}) can be reformulated as $dim{\mathcal M}=n^2$, where $\mathcal M=\{x:\ xU=Ux,\ U\in {\mathcal U}\}$ is the algebra of bounded operators commuting with elements of the group $\mathcal U$.} 

Suppose that $\mathcal T$ is an Abelian subgroup of $\mathcal U$. Let us define a marginal probability distribution on the set of all left cosets $[U]=\{UT,\ T\in {\mathcal T}\}$ as follows
\begin{equation}\label{marg}
p_{[U]}=\sum \limits _{T\in {\mathcal T}}\pi _{UT}.
\end{equation}
We shall say that the majorization property takes place if
\begin{equation}\label{major}
p_{[U]}\le p_{[V]}\ \Rightarrow \ \pi _{UT}\le \pi _{VS}\ \text{for all}\ S,T\in {\mathcal T}.
\end{equation}

At the moment the marginal distribution $(p_{[U]})$ is indexed by elements of cosets $[U]$. We can 
pick up $\frac {|\mathcal U|}{|\mathcal T|}$ indexes $j$ and $U_j\in {\mathcal U}$ such that cosets $[U_j]\neq [U_k]$ for $j\neq k$. Moreover, $p_{[U_j]}\le p_{[U_k]}$ if $j\ge k$. Denote $P=\{p_j=p_{[U_j]},\ 1\le j\le \frac {|\mathcal {U}|}{|\mathcal {T}|}\}$ the corresponding probability distribution.

{\bf Proposition 2.} {\it Suppose that $\mathcal A$ has a codimension $n$, $|\mathcal {U}|\le n|\mathcal {T}|$ and the majorization condition (\ref {major}) holds true. Then,
$$
\Phi \prec P.
$$
}
{\bf Remark 2.} {\it The condition $\Phi \prec P$ doesn't imply that the eigenvalues of $\Phi (\ket {f}\bra {f})$ form a probability distribution majorized by $P$ in the sense of Karamata for any choice of a unit vector $f\in H$. In some partial cases this estimate can be useless. The condition $dimH=n$ can be satisfied for the case in which $\Phi $ can be represented in the form of a composition of several quantum channels like it appears in the next section.}

Proof.

Take a unit vector $f\in H$. Pick up eigenvectors $(e_j)$ corresponding to eigenvalues $(\lambda _j^f)$ of $\Phi (\ket {f}\bra {f})$ numbered in the descending order $\lambda _j^f=\lambda _j^{f\downarrow}$. Notice that the indexes $j$ were picked up in such a way that if $j>k$ than $\lambda _j\le \lambda _k$ and $p_j\le p_k$. Moreover, since the codimension equals $n$ we obtain that a number of different eigenvectors $e_j$
is not less than $dimH\ge n\ge \frac {|\mathcal {U}|}{|\mathcal {T}|}$. 
Then,
\begin{equation}\label{sum}
\sum \limits _{j=1}^k\lambda _j^f=\sum \limits _{j=1}^k\sum \limits _{U\in {\mathcal U}}\pi _U\braket {e_j,U\ket {f}\bra {f}U^*e_j}=
\sum \limits _{j=1}^k\sum \limits _{U\in {\mathcal U}}\pi _U|\braket {e_j,Uf}|^2
\end{equation}
Since ${\mathbb E}(\ket {f}\bra {f})\cong \frac {1}{n}{\rm I}_n\otimes \sigma $ and $Tr(\sigma )=1$ due to (\ref {dim}) we get
\begin{equation}\label{inequality}
||{\mathbb E}(\ket {f}\bra {f})||\le \frac {1}{n}.
\end{equation}
Taking into account (\ref {exp}) with $|{\mathcal U}|\le n|\mathcal {T}|$ we obtain
$$
\sum \limits _{U\in {\mathcal U}}\braket {e_j,U\ket {f}\bra {f}U^*e_j}=\sum \limits _{U\in {\mathcal U}}|\braket {e_j,Uf}|^2=|{\mathcal U}|\braket {e_j,{\mathbb E}(\ket {f}\bra {f})e_j}\le |\mathcal {T}|
$$
in virtue of (\ref {inequality}).

Put $\alpha _{j,U}=|\braket {e_j,Uf}|^2$, then $0\le \alpha _{j,U}\le 1$.
Now we need to estimate the sum $\sum \limits_{j=1}^k\sum \limits _{U\in {\mathcal U}}\pi _U\alpha _{j,U}$ under the
conditions 
$$
\mathlarger {\sum }_{\substack {U\in {\mathcal U}_0\subset {\mathcal U},\\ |{\mathcal U}_0|=k|\mathcal {T}|}}\pi _U\le \sum \limits _{j=1}^kp_j
\ \text{and}\ \sum \limits _{U\in {\mathcal U}}\alpha _{j,U}\le |\mathcal {T}|.
$$
It results in (\ref {sum}) is less or equal to the sum $\sum \limits _{j=1}^kp_j$ completed so far $k|\mathcal {T}|\le |\mathcal U|$, where $p_j\in P$ form the marginal probability distribution introduced by (\ref {marg}) and (\ref {major}).

To complete the proof we need to show that for any choice of orthogonal unit vectors $\tilde e_j,\ 1\le j\le k$ we have
$$
\sum \limits _{j=1}^k\braket {\tilde e_j,\Phi (\ket {f}\bra {f})\tilde e_j}\le \sum \limits _{j=1}^k\lambda _j^f,\ k|\mathcal {T}|\le |\mathcal U|,
$$
where $(\lambda _j^f)$ are eigenvalues of $\Phi (\ket {f}\bra {f})$ numbered in the descending order.
But it follows from the minimax properties of eigenvalues.

$\Box $

\section {Exploiting the extension of Abelian groups}

Consider a finite Abelian group $S,\ |S|=n$. Since $S$ is a direct product of cyclic groups ${\mathbb Z}_{p_j}$ it is generated by commuting elements 
$$
s_j,\  s_j^{p_j}=e,\ 1\le j\le m,\ \prod _{j=1}^mp_j=n.
$$ 
Denote ${\mathbb K}$ the set of all multi-indeces $K=(k_1,\dots ,k_m),\ k_j\in {\mathbb Z}_{p_j}$. Fix the orthonormal basis $\{e_K,\ K\in {\mathbb K}\}$ in a Hilbert space $H,\ dimH=n$. Then one can define unitary operators $U_j$ and $T_j$ in $H$ by the formula
$$
U_je_K=e_{k_1\dots k_{j-1}\ k_js_j\ k_{j+1}\dots k_m},
$$
$$
T_je_K=e^{i\frac {2\pi k_j}{p_j}}e_K
$$
Let ${\mathcal U}$ be the unitary group generated by $\{U_j,T_j,\ 1\le j\le m\}$.
Denote $\mathcal T$ the subgroup of $\mathcal U$ generated by $\{T_j,\ 1\le j\le m\}$.

{\bf Proposition 3.} {\it The group $T$ is an Abelian normal subgroup of $\mathcal U$. Moreover, $\mathcal U/T$ is isomorphic to the subgroup ${\mathcal U}_0$ of $\mathcal U$ and we have the short exact sequence
\begin{equation}\label{exact}
I\to {\mathcal T}\rightarrow {\mathcal U}\rightarrow {\mathcal U/T}\rightarrow I
\end{equation}
}

Proof.

The group $\mathcal T$ is Abelian by a construction. This subgroup is normal because $U_jT_jU_j^*=e^{i\frac {2\pi s_j}{p_j}}T_j$ and
$U_jT_kU_j^*=T_k$ for $j\neq k$. A coset $[U_j]$ consists an element $U_j$ acting on $e_K$ without multiplying on any phase factor.
Finally the maps in (\ref {exact}) are defined by ${\mathcal T}\ni T\to T\in {\mathcal U}\to [I]\in {\mathcal U/T}$.

$\Box $

{\bf Remark 3.} {\it It follows from (\ref {exact}) that
the group $\mathcal U$ can be considered as the extension of the Abelian group ${\mathcal U}_0$ by means of the Abelian group $\mathcal T$. The other way to look at $\mathcal U$ is to define the action $\alpha $ of ${\mathcal U}_0$ on $\mathcal T$ by the formula
$$
\alpha _U(T)=UTU^*,\ U\in {\mathcal U}_0,\ T\in {\mathcal T}.
$$
In the case,
$$
{\mathcal U}\cong {\mathcal T}\rtimes _{\alpha }{\mathcal U}_0.
$$
}

{\bf Proposition 4.} {\it For the expectation (\ref {exp}) we get
\begin{equation}\label{expect}
{\mathbb E}(\rho )=\frac {1}{n}{\rm I}_H,\ \rho \in \mathfrak {S}(H).
\end{equation} 
Thus, the algebra $\mathcal A$ has a codimension $n$ (\ref {dim}).}

Proof.

We need to show that the commutant ${\mathcal U}'$ is trivial. In order to make sure of this, it is enough to note that the group $\mathcal U$ is a direct product of irreducible unitary representations of Heisenberg-Weyl groups generated by operators $U_j,T_j,\ 1\le j\le m$.

$\Box $

Consider the quantum channel (\ref {kanal}) generated by some probability distribution on our group $\mathcal U$.

{\bf Theorem 1.} {\it Let $F$ be an arbitrary convex function on $[0,1]$. Suppose that the majorization condition (\ref {major}) holds true. Then,
\begin{equation}\label{CO}
\inf \limits _{\rho \in \mathfrak {S}(H)}Tr(F(\Phi (\rho )))=\sum \limits _jF(p_j).
\end{equation}
}

Proof.

The infimum in (\ref {CO}) is achieved on some pure state $\ket {f}\bra{f}$. Denote $\lambda _j^f$ the eigenvaules of $\Phi (\ket {f}\bra {f})$ numbered in the descending order. 
Applying the expectation (\ref {expect}) to $\ket {f}\bra {f}$ we obtain
$$
{\mathbb E}(\ket {f}\bra {f})=\frac {1}{n}{\rm I}_n
$$
due to Proposition 4, where $n=dimH$. Moreover, the number of cosets $|{\mathcal U/T}|=|{\mathcal U}_0|=n$. Hence the channel $\Phi$ satisfies the conditions of Proposition 2 and
$$
\sum \limits _{j=1}^k\lambda _j^f\le \sum \limits _{j=1}^kp_j,\ 1\le k\le n. 
$$

Applying the Karamata's theorem we obtain
\begin{equation}\label{ineq}
\sum \limits _{j=1}^nF(\lambda _j^f)\le \sum \limits _{j=1}^nF(p_j).
\end{equation} 
Take a unit vector $f=e_K$,  then
$$
\lambda _j^{f}=p_j
$$
for any choice of the multi-index $K$.
So the equality in (\ref {ineq}) is achieved on $f=e_K$.
Finally, notice that
$$
\sum \limits _{j=1}^nF(\lambda _j^f)=Tr(F(\Phi (\ket {f}\bra {f}))).
$$

$\Box $

Pick up $M$ unitary groups ${\mathcal U}_j$ obtained by means of the procedure described above acting in Hilbert spaces $H_j,\ dimH_j=n_j.$ Put 
\begin{equation}\label{gruppa}
{\mathcal U}=\bigotimes _{j=1}^M{\mathcal U}_j,\ H=\bigotimes _{j=1}^MH_j.
\end{equation}
 Denote $P^{(j)}=\{p_k^{(j)},\ 1\le k\le n_j\}$ the marginal probability distributions (\ref {marg}) each numbered in the descending order. Consider the channels $\Phi _j$ determined by (\ref {kanal}) with the group ${\mathcal U}_j$. Put $n=\prod _{j=1}^Mn_j$.
Then, the channel $\Phi $ determined by (\ref {kanal}) for the case of the group (\ref {gruppa}) can be represented as
$$
\Phi =\bigotimes _{j=1}^M\Phi _j.
$$

{\bf Theorem 2.} {\it Let $F$ be an arbitrary convex function and the majorization condition (\ref {major}) be satisfied for each $P^{(j)}$, $1\le j\le M$.
Then
\begin{equation}\label{equ}
\inf \limits _{\rho \in \mathfrak {S}(H)}Tr(F(\Phi (\rho )))=\sum \limits _{j_1=1}^{n_1}\dots \sum \limits _{j_M=1}^{n_M}
F\left (\prod _{k=1}^Mp_{j_k}^{(k)}\right ),
\end{equation}
}

Proof.

Consider channels
$$
\tilde \Phi _j={\rm Id}_{\otimes _{k=1}^{j-1}H_j}\otimes \Phi _j\otimes {\rm Id}_{\otimes _{k=j+1}^MH_k},\ 1\le j\le M,
$$
acting non-identically only on operators of $\mathfrak {S}(H_j)$.
Arguing in the same way as under Proof of Theorem 1 we obtain that the algebra ${\mathcal A}_j$ consisting of fixed elements with respect to the action of ${\mathcal U}_j$  has a codimension $n_j$. Thus, the expectation ${\mathbb E}_j$ on $\mathcal {A}_j$
has the form
$$
{\mathbb E}_j(\ket {f}\bra {f})\cong \frac {1}{n_j}{\rm I}_{n_j}\otimes \sigma ,\ Tr(\sigma )=1.
$$
Applying Proposition 2 we obtain $\tilde \Phi _j\prec P^{(j)}$ for each $j,\ 1\le j\le M$. Hence, Proposition 1 implies that 
\begin{equation}\label{equ2}
\Phi =\otimes _{j=1}^M\Phi _j=\tilde \Phi _1\circ \tilde \Phi _2\circ \dots \circ \tilde \Phi _M\prec P^{(1)}\star P^{(2)}\star \dots \star P^{(M)}.
\end{equation}
Take a unit vector $f\in H$ and consider the eigenvalues $\lambda _s^f$ of $\Phi (\ket {f}\bra {f})$ numbered in the descending order.  It follows from (\ref {equ2}) that
$$
\sum \limits _{s=1}^k\lambda _s^f\le r_s,\ 1\le s\le n,
$$
where $r_s=r_s^{\downarrow}\in P^{(1)}\star P^{(2)}\star \dots \star P^{(M)}$. Applying the Karamata's theorem we obtain (\ref {equ}) with inequality "$\le $". Substituting a factorizable projection $\rho =\ket {f}\bra {f}$ we obtain the equality.

$\Box $

{\bf Corollary.} {\it When the conditions of Theorem 2 are fulfilled we get
$$
||\otimes _{j=1}^M\Phi _j||_p=\prod _{j=1}^M||\Phi _j||_p=\prod _{j=1}^M\left (\sum \limits _{k_j=1}^{n_j}(p_{k_j}^{(j)})^p\right )^{\frac {1}{p}},\ p>1,
$$
$$
S_{min}(\Phi)=\sum \limits _{j=1}^MS_{min}(\Phi _j)=-\sum \limits _{j=1}^M\sum \limits _{k=1}^{n_j}p_k^{(j)}\log p_k^{(j)},
$$
\begin{equation}\label{FORCOR}
C(\Phi )=\sum \limits _{j=1}^MC(\Phi _j)=\log(n)+\sum \limits _{j=1}^M\sum \limits _{k=1}^{n_j}p_k^{(j)}\log p_k^{(j)}.
\end{equation}
}

Proof.

To obtain two first equalities it suffices to apply Theorem 2 for convex functions $F(t)=t^p$ and $F(t)=t\log t$. The last one can be derived from the inequality
$$
C(\Phi )\le \log dimH-S_{min}(\Phi )
$$
in which we have the equality if the minimum of entropy is achieved for the ensemble of states forming the resolution of identity. For our case this claim is fulfilled. 

$\Box $

\section{The connection with Boson quantum channels}

Fiber-optic communications  are used in the actual transmission of information \cite{Kirsanov}. In this case, the main problem in this direction is the exponential decay of the transmitted quantum signal with the distance. The technological approach enables us to realize quantum states’ repetition by optical amplifiers keeping states’ wave properties and phase coherence. We consider amplifiers as a partial case of Boson quantum channels. Since different amplifiers are statistically independent the distance of transmission is one-to-one correspondence with the composition of the channels' action. Here we will give only the simplest model that allows us to move from Boson channels to qubit channels.

Following to \cite {Holevo} each one-mode Boson amplifier $\Psi $ is described by the action on the characteristic function $F_{\rho }(x,y)=Tr(\rho e^{ixQ+iyP})$ of a quantum state $\rho $ having the form
\begin{equation}\label{Boson}
F_{\Psi (\rho )}(x,y)=e^{-\frac {1}{2}(\frac {k-1}{2}+N_c)(x^2+y^2)}F_{\rho }(\sqrt kx,\sqrt ky), 
\end{equation}
where $k>1$ is a coefficient of amplification and $(Q,P)$ are the standard position and momentum operators and $N_c$ is a channel noise. Since the classical capacity $C(\Psi )$ of $\Psi $ coincides with the one-shot capacity \cite{GHG} the formula for the classical capacity is given by \cite{Holevo}
$$
C(\Psi )=g(kN+N_c)-g(N_c),
$$ 
where $g(x)=(x+1)\log (x+1)-x\log x ,\ x>0,\ g(0)=0$. Here we suppose that the restriction 
\begin{equation}\label{restrict}
Tr\left (\overline {\rho} \frac {Q^2+P^2}{2}\right )\le \frac {1}{2}+N
\end{equation} 
is imposed on the average quantum states 
$\overline {\rho}$ obtaining during the transmission. 
It is known that \cite{Holevo, GHG} the classical capacity is attained on coherent states $\ket {z}\bra {z},\ z\in {\mathbb C},$ such that
$$
C(\Psi)=S\left (\Psi \left (\frac {1}{\sqrt {\pi N}}\int \limits _{\mathbb {C}}e^{-\frac {|z|^2}{N}}\ket {z}\bra {z}d^2z\right )\right )-S(\Psi (\ket {z}\bra {z})).
$$
In practice we use only a finite number of coherent states to encode the information with respect to the discrete probability distribution $\pi $ on $\mathbb C$ satisfying (\ref {restrict}). Thus, we obtain a formula having the form (\ref {FORCOR}), where
$$
S\left (\Psi \left (\int \limits _{\mathbb {C}}\ket {z}\bra {z}d\pi (z)\right )\right )\equiv \log n
$$
and
$$
S(\Psi (\ket {z}\bra {z}))\equiv -\sum \limits _{j=1}^M\sum \limits _{k=1}^{n_j}p_k^{(j)}\log p_k^{(j)}.
$$
To achieve a complete analogy with (\ref {FORCOR}) we would need to fulfill the condition $\Psi (\overline {\rho })=\overline {\rho }$ for
$$
\overline {\rho}=\int \limits _{\mathbb {C}}\ket {z}\bra {z}d\pi (z).
$$
Moreover, $\overline \rho $ should be a projection $\overline {\rho}^2=\overline \rho $ that is not fair in general \cite {Alekseev}. We plan to study this question separately.

\section{Conclusion}

We continue to develop the theory, the foundations of which were laid in \cite{Amo, Amo2}. A thorough analysis of the conditions that allow us to prove that the use of entangled states does not give a gain for the class of channels under consideration has shown that the operation of taking a tensor product can be replaced by a composition of channels of the class under consideration. Such operations of decomposition naturally appear in quantum computing schemes, which gives hope for further advances in this direction in the future. In addition to calculating the classical capacity of channels directly, it was possible to find a uniform approach to calculating a wide class of characteristics being convex functions of the eigenvalues of the output states of the channels. We also discuss the connection of our approach with the informational characteristics of Bosonic quantum channels.

\section*{Acknowledgments} The author is grateful to V.I. Yashin for stimulating discussions. 




\end{document}